\documentclass[twocolumn]{aastex63}

\usepackage{amssymb}
\usepackage{amsmath}
\usepackage{txfonts}

\usepackage{natbib}
\hypersetup{linkcolor=red,citecolor=blue,urlcolor=cyan}

\usepackage{graphicx}
\usepackage{dcolumn}
\usepackage{bm}
\usepackage{color}
\usepackage{soul}
\usepackage{mathtools}
\usepackage[T1]{fontenc}

\def\aap{A\&A}

\def\apjl{ApJL}
\def\apjs{ApJS}

\date{February 15th, 2021}

\shorttitle{Probing elastic quark phases in hybrid stars with radius measurements}
\shortauthors{Pereira, Bejger, Tonetto, Lugones, Haensel, Zdunik, and Sieniawska}
\graphicspath{{./}{figures/}}

\begin{document}

\title{Probing elastic quark phases in hybrid stars with radius measurements}

\correspondingauthor{Jonas P.~Pereira}
\email{jpereira@camk.edu.pl}

\author{Jonas P.~Pereira}
\affiliation{Nicolaus Copernicus Astronomical Center, Polish Academy of Sciences, Bartycka 18, 00-716, Warsaw, Poland}

\author{Micha{\l} Bejger}
\affiliation{Nicolaus Copernicus Astronomical Center, Polish Academy of Sciences, Bartycka 18, 00-716, Warsaw, Poland}

\author{Lucas Tonetto}
\affiliation{Dipartimento di Fisica, ``Sapienza'' Universit\`a di Roma \& Sezione INFN Roma1, P.A. Moro 5, 00185, Roma, Italy}
\affiliation{Universidade Federal do ABC, Centro de Ci\^encias Naturais e Humanas, Avenida dos Estados 5001- Bang\'u, CEP 09210-580, Santo Andr\'e, SP, Brazil}

\author{Germ\'an Lugones}
\affiliation{Universidade Federal do ABC, Centro de Ci\^encias Naturais e Humanas, Avenida dos Estados 5001- Bang\'u, CEP 09210-580, Santo Andr\'e, SP, Brazil}

\author{Pawe{\l} Haensel}
\affiliation{Nicolaus Copernicus Astronomical Center, Polish Academy of Sciences, Bartycka 18, 00-716, Warsaw, Poland}

\author{Julian Leszek Zdunik}
\affiliation{Nicolaus Copernicus Astronomical Center, Polish Academy of Sciences, Bartycka 18, 00-716, Warsaw, Poland}

\author{Magdalena Sieniawska}
\affiliation{Astronomical Observatory, University of Warsaw, Al. Ujazdowskie 4, 00-478 Warsaw, Poland}

\begin{abstract}
The internal composition of neutron stars is currently largely unknown. Due to the possibility of phase transitions in quantum chromodynamics, stars could be hybrid and have quark cores. 
We investigate some imprints of elastic quark phases (only when perturbed) on the dynamical stability of hybrid stars.
We show that they increase the dynamical stability window of hybrid stars in the sense that the onset of instabilities happen at larger central densities than the ones for maximum masses. 
In particular, when the shear modulus of a crystalline quark phase is taken at face value, the relative radius differences between elastic and perfect-fluid hybrid stars with null radial frequencies (onset of instability) would be up to $1-2\%$. Roughly, this would imply a maximum relative radius dispersion (on top of the perfect-fluid predictions) of $2-4\%$ for stars in a given mass range {\it exclusively} due to the elasticity of the quark phase.
In the more agnostic approach where the estimates for the quark shear modulus only suggest its possible order of magnitude (due to the many approximations taken in its calculation), the relative radius dispersion uniquely due to a quark phase elasticity might be as large as $5-10\%$. 
Finally, we discuss possible implications of the above dispersion of radii for the constraint of the elasticity of a quark phase with electromagnetic missions such as NICER, eXTP and ATHENA.
\end{abstract}


\section{Introduction}

Neutron stars (NSs) are very compact remnants of stellar evolution that offer ways to probe particle and dense-matter physics aspects not possible in terrestrial laboratories. So far, the constitution and internal structure of an NS is not known in detail. The observational evidence for NSs with masses around $2\,M_\odot$ \citep{2010Natur.467.1081D,2013Sci...340..448A,2016ApJ...832..167F,2020NatAs...4...72C} has helped constrain some microphysical models, but there are still many viable possibilities. The advent of the multi-messenger astronomy, thanks to direct detections of gravitational waves (GWs) by the Advanced LIGO \citep{ligosc2015} and Advanced Virgo \citep{Acernese_2014} detectors' network, promises further progress. Specifically, the observation of the GW170817 event, undoubtedly an inspiral of a binary NS system by the LIGO-Virgo collaboration \citep{2017PhRvL.119p1101A,2018arXiv180511581T,2019PhRvX...9a1001A}, has provided the first measurement of the late inspiral mutual tidal deformability of the components (in the form of the component mass-weighted sum of individual tidal deformabilities). This value, albeit burdened by large measurement errors, already constrains some microphysical models for the dense matter equation of state (EOS) with and without phase transitions (see, e.g., \cite{2020arXiv200603168C,2020arXiv200607194M,2020arXiv200603789B,2020arXiv200600839M,2020arXiv200510543F} and references therein).
However, this single observation is not sufficient to definitely answer the question whether the GW170817 NS components were one-phase (hadronic) stars or exhibited phase transitions \citep{2018PhRvL.121i1102D,2020PhRvD.101f3007E,2020Univ....6...81B}. Additional inspiral and merger GW observations \citep{2020PhRvD.101f3022L,2020PhRvD.101d4019C}, as well as the post-merger \citep{2019PhRvL.122f1102B,2019PhRvL.122f1101M,2020PhRvL.124q1103W} observations might resolve this ambiguity. It is of general interest to combine these future measurements with other observable quantities which could differentiate purely hadronic stars from stars with quark cores (hybrid stars). 

The same goal may be achieved with radius measurements by means of electromagnetic observations; note that GW measurements of binary NS inspirals do not provide direct radius measurements. This might now be possible with the NICER mission \citep{2016ApJ...832...92O}, which has already constrained an NS radius with an uncertainty smaller than $10\%$ (at 1-sigma confidence level) \citep[see][]{Bilous_2019,Riley_2019,Raaijmakers_2019,Miller_2019,Bogdanov_2019,Bogdanov_2019_,Guillot_2019}. Future missions such as the eXTP \citep{2016SPIE.9905E..1QZ} and ATHENA \citep{2020ApJ...888..123M} promise even smaller uncertainties to NS radii (a few percent).\footnote{Such constraints are not direct but rather byproducts of NS emission models, which attempt to explain for instance their light-curves and pulse profiles \citep{2016ApJ...832...92O,Riley_2019,Miller_2019,2020ApJ...889..165D}.} An important question is whether radius differences of stars with and without phase transitions could be larger than NICER and eXTP uncertainties, and if there are suitable observational candidates presenting such features. When compared to purely hadronic stars, some models for the ``third family'' of NSs \citep{2017PhRvC..96d5809A,2017PhRvL.119p1104A,2019arXiv191209809C,2019PhRvC.100b5802M} represent such a theoretical possibility. They are hybrid stars exhibiting substantial matter softening (resulting in a local minimum in the mass-radius sequence), for instance due to a first-order phase transition, where the density at the quark-hadron interface is discontinuous \citep{2008A&A...479..515Z}.
So far---due to the lack of a quark phase EOS---, there is not a definite threshold mass for hybrid stars, although there are statistical suggestions they might be large \citep{Annala:2019puf}. Any possibility could be observationally covered since there is evidence that the NS mass function supports a bimodal distribution centered at approximately $1.4\,M_{\odot}$ and $1.8\, M_{\odot}$, with spread around $0.2\,M_{\odot}$ \citep{2018MNRAS.478.1377A}. In summary, one has interesting candidates and detectors to probe aspects pertaining to hybrid stars.

However, in spite of the above optimistic observational scenario, ambiguities still remain concerning other aspects of their quark phases. 
It has been recently shown that the quark phase relevant to NSs could also be crystalline \citep{2006PhRvD..74i4019R}, with a shear modulus possibly up to a thousand times larger than the one of usual crusts \citep{2007PhRvD..76g4026M}. The question arises how it would be possible to differentiate between an elastic and a perfect-fluid model for a hybrid star when the quark phase is concerned.\footnote{The elasticity of the crust of an NS is expected to change negligibly its dynamical stability because it contributes to a mass around $1\%$ of the stellar mass.} This would be of huge interest because it might allow for constraints of the quark phase. Tidal deformations seem excellent observables \citep{2017PhRvD..95j1302L,2019PhRvD..99b3018L} in this regard, although only third generation GW detectors, many combined GW measurements of the Advanced detectors, or an (unlikely) nearby high signal-to-noise binary NS merger might reach the required precision; see, e.g., \citet{2020arXiv200310781P} and references therein.

Electromagnetic observations may provide a similar and complementary information in the near future if models of stable elastic and perfect-fluid stars exhibit radii differences larger than the NICER and eXTP uncertainties. For a given EOS, radial stability analyses could easily determine this. However, one would not expect that null eigenfrequencies in elastic  hybrid stars should satisfy the usual dynamical stability rule \citep{Wheeler1965,1988ApJ...325..722F,2011MNRAS.416L...1T}
$\partial M/\partial \rho_c= 0$, where $M$ is the background mass of the star and $\rho_{\rm{c}}$ is its central density, though this is the case for elastic stars without phase transitions \citep{2004CQGra..21.1559K}. Indeed, $\partial M/\partial \rho_c\geq 0$ for real radial eigenfrequencies arises in the context of cold, catalyzed matter constituting a perfect fluid \citep{Wheeler1965,1988ApJ...325..722F,2011MNRAS.416L...1T}, or a fully elastic star \citep{2004CQGra..21.1559K}. For hybrid stars with elastic and liquid components, perturbation analyses are important in order to come up with the minimum radii of stable systems. 

The article is arranged as follows. In Sec.~\ref{solid stars}, we lay out the models for elastic hybrid stars we will work with. Section~\ref{model-radial-perturbations-solid-stars} is devoted to the deduction of the field equations for radial perturbations in elastic hybrid stars and the appropriate matching conditions to be taken in this case. In Sec.~\ref{results} we present our main results; we mainly show that an elastic quark phase increases the stability window of a hybrid star. Finally, we discuss all points raised in Sec.~\ref{discussion}. Unless otherwise stated, we work with geometric units and our metric signature convention is $(-,+,+,+)$.

\section{Elastic hybrid stars}
\label{solid stars}

In this work we assume that a hybrid star is constituted of a quark phase and a hadronic phase and that they are split by a sharp phase-transition surface. By this we mean that we do not consider here a mixed phase and that the density at the phase-splitting surface is in general discontinuous.

\subsection{Models for hybrid stars}
\label{models_hybrid_stars}
\subsubsection{Chiral effective model + MIT bag-like}
\label{quark_phase_model} 

In recent years, the chiral effective field theory (cEFT) has provided a framework that allows a systematic expansion of nuclear forces at low energies based on the symmetries of quantum chromodynamics (QCD) \citep{Epelbaum:2008ga,Hammer:2012id}. The cEFT interactions can be employed in microscopic many-body frameworks in order to derive an EOS for neutron-rich matter. Since cEFT is an effective low-energy theory, it contains a breakdown scale that imposes an upper density limit  in the calculation of the EOS. A  conservative choice of  $1.1 \, \rho_{\rm{sat}}$  ($\rho_{\rm{sat}}  =\mathrm{2.7\times10^{14} g\,cm^{-3}}$) was adopted in \cite{Hebeler:2013nza}.  Above this density, the EOS can be extended by means of piecewise polytropic EOSs which take into account the validity of causality and the consistency with observed $2 M_\odot$ pulsars. Here we adopt the model of \cite{Hebeler:2013nza} which employs a set of three polytropes  valid in three consecutive density regions. This procedure leads to a very large number of EOSs, which verify the physical and observational constraints mentioned before. For the use in astrophysical applications,  \cite{Hebeler:2013nza}  provides detailed numerical tables for three representative EOS labeled as \textit{soft}, \textit{intermediate} and \textit{stiff}. For densities below $0.5 \, \rho_{\rm{sat}}$ the BPS crust EOS is used \citep{Baym:1971pw,1973NuPhA.207..298N}.

For the quark (innermost) phase, we take a phenomenological microscopic model put forth by \cite{2005ApJ...629..969A} which leads to the following EOS (see \cite{2018ApJ...860...12P} for further details):
\begin{eqnarray}
p(\rho) &=& \tfrac{1}{3}(\rho - 4B) \nonumber\\
&-&\frac{a_2^2}{12\pi^2a_4}\left[1+\sqrt{1+\frac{16\pi^2a_4}{a_2^2}(\rho -B)} \right],\label{eos_qm}
\end{eqnarray}
where $(B,a_4,a_2)$ are free parameters. Here, $B$ emulates the quantum chromodynamics confinement; $a_4$ accounts for the quark strong interactions and $a_2$ encompasses aspects such as the strange quark mass and color superconductivity \citep{2005ApJ...629..969A}. Further microscopic details on this model can be found in \cite{2018ApJ...860...12P}.
These EOSs are glued together through Gibbs criteria for phases in mechanical, thermal and chemical equilibrium (oftentimes called ``the Maxwell construction'') \citep{1986bhwd.book.....S}. They lead to a unique phase transition pressure $p_{\rm pt}$ and density jump at the phase-splitting surface. 

\subsubsection{SLy4 + effective polytropic + simple MIT bag}
\label{sly4}
The SLy4 EOS is based on Skyrme-type effective interactions in nuclear matter, 
originated in the ideas presented in the seminal  works of Skyrme \citep{1956PMag....1.1043S,1958NucPh...9..615S}. 
The Skyrme-type
effective interactions were  first successfuly applied, within the density functional theory (DFT), to  atomic nuclei \citep{1972PhRvC...5..626V}
as well as to the  nuclei in NS crusts  \citep{1973NuPhA.207..298N}.
The effective interaction contains two-body terms and terms resulting from the 
averaging of the three-body interaction. The SLy effective hamiltonians \citep{1997NuPhA.627..710C,1998NuPhA.635..231C}
contain a number of parameters that have been adjusted to  reproduce experimental 
data on selected neutron rich atomic nuclei. A specific  SLy4 effective hamiltonian is
consistent  with  many-body calculations  of the EOS of pure neutron matter with realistic two-nucleon and three-nucleon force (UV14+UVII model of \cite{1988PhRvC..38.1010W}),
and so is suitable to describe the NS core. A  unified  SLy4 EOS for NS  was obtained,  based on SLy4 effective interaction model. It describes in a physically consistent way  (i.e. starting from the same nuclear effective interaction)  the structure and EOS of the crust and the core, including the transition between them \citep{2001A&A...380..151D}.

In order to construct equations of state for the hadronic part, we connect the SLy4 EOS to a relativistic polytrope $p=\kappa_{\rm ef} n_b^{\gamma}$ [$\rho = p/(\gamma -1) +  n_b m_b$] at a baryon density $n_0$ and extend it up to a higher baryon density $n_1$. In the above expressions, $n_b$ is the baryon density, ($n_0$, $n_1$, $\gamma$) are free parameters, and $m_b$ (baryon mass) and $\kappa_{\rm ef}$ are adjustable parameters assuring the continuity of the pressure and the chemical potential at $n_0$. The simple MIT bag EOS $p=c_s^2(\rho-\rho_{\star})$ describing the quark phase is connected to the above hadronic EOS in particular. In this work, we choose $\gamma=4.5$ and $n_1=0.335$ fm$^{-3}$ (above 2$\rho_{\rm sat}$), and take $c_s^2=1$ in order to also probe aspects of stiff quark matter. The baryon density jump at the quark-hadron interface is a free parameter and chemical and mechanical equilibrium there imply a unique density jump and $\rho_{\star}$. For further details on this model, see \cite{2019A&A...622A.174S,2020arXiv200310781P}.

\subsection{Shear moduli}

\subsubsection{Shear modulus for the hadronic phase}
\label{hadronic_phase}
Here we take into account crystalline aspects expected to the crust of NSs  \citep{2008LRR....11...10C}, obtained from known constraints to nuclear matter. We work with a simple linear model for the shear modulus \citep{2011PhRvD..84j3006P}, namely 
\begin{equation}
    \tilde{\mu}_{h}= \tilde{\mu}_0+\kappa p\label{mu-had},
\end{equation}
where $\tilde{\mu}_0$ and $\kappa$ are tunable constants, and $p$ is the pressure. We take $\tilde{\mu}_0\simeq 0$ and $\kappa = 0.015$ \citep{2008LRR....11...10C,2020arXiv200310781P}. We assume that the onset of the crust elasticity is at $2\times 10^{14}$~g~cm$^{-3}$ and that it finishes at $10^{7}$~g~cm$^{-3}$, where the liquid ocean/envelope is supposed to start \citep{2020arXiv200310781P}. It is already known that such shear modulus leads to negligible tidal deformation changes in ordinary NSs \citep{2011PhRvD..84j3006P,2020PhRvD.101j3025G}, described in terms of zero-frequency nonradial perturbations. Therefore, one would expect that for other types of perturbations they would also lead to negligible changes when compared to the perfect-fluid results. For completeness, we nevertheless include them in our analysis.

\subsubsection{Shear modulus for the quark phase}
Following the model of \cite{2007PhRvD..76g4026M}, we investigate crystalline aspects of color superconducting quark matter (the LOFF phase) \citep{2008RvMP...80.1455A}, whose shear modulus can be hundreds of times larger than the shear modulus of the crust for some crystal structures \citep{2007PhRvD..76g4026M}. Under certain assumptions, it has been shown that the shear modulus of the LOFF phase is \citep{2007PhRvD..76g4026M}
\begin{equation}
    \tilde{\mu}_{q}\simeq 2.5 \left(\frac{\Delta}{10\, \mbox{MeV}} \right)^2\left(\frac{\mu_q}{400\, \mbox{MeV}} \right)^2 \mbox{MeV fm$^{-3}$}\label{mu-quark},
\end{equation}
where $\mu_q$ is the quark chemical potential and $\Delta$ is the crystalline pairing gap parameter. Conservative values for $\Delta$ are in the range $5-25$ MeV; $\mu_q$ typically varies from $350$ to $500$ MeV \citep{2007PhRvD..76g4026M}. 
Therefore, taken at face value, the shear modulus of the crystalline quark phase should be in the interval 
\begin{equation}
    0.50\, \mbox{MeV fm$^{-3}$} \lesssim \tilde{\mu}_q \lesssim 25\, \mbox{MeV fm$^{-3}$}.
\end{equation}
However, the pre-factor of Eq.~\eqref{mu-quark} might vary considerably due to the many approximations and crystalline structures assumed for its calculation \citep{2014RvMP...86..509A,2014PhRvD..89j3014M}. This means that the above range of $\tilde{\mu}_q$ might also change appreciably. Due to this, we take the agnostic point of view that the estimates of \cite{2007PhRvD..76g4026M} only point to the correct order of magnitude of the crystalline quark shear modulus. Thus, we allow the shear modulus of the quark phase (taken at the fiducial chemical potential 400 MeV) to be as large as $50-60$ MeV fm$^{-3}$. In order to encompass the above uncertainty, hereafter we replace Eq. \eqref{mu-quark} by the phenomenological shear modulus
\begin{equation}
    \tilde{\mu}_{q, \mu_q}^{(p)}= 2.5 \alpha \left(\frac{\Delta}{10\, \mbox{MeV}} \right)^2\left(\frac{\mu_q}{400\, \mbox{MeV}} \right)^2 \mbox{MeV fm$^{-3}$}\label{mu-quark-p},
\end{equation}
where $1\lesssim \alpha \lesssim 4$. 
In our numerical analysis, we take $\mu_q$ as directly coming from the assumed microphysics.

\section{Radial perturbations in elastic neutron stars}
\label{model-radial-perturbations-solid-stars}

The issue of linear radial perturbations in elastic stars has been addressed for the first time in \cite{2004CQGra..21.1559K}. They have shown that for one-phase stars the ordinary dynamical stability rules are not affected by elastic aspects of matter. However, to the best of our knowledge, this issue has not been addressed for hybrid stars with elastic and liquid parts in the presence of phase conversions \citep{2018ApJ...860...12P}. 

\subsection{Field equations for radial perturbations in perfect fluids}

The equations describing the dynamics of perturbations have been obtained in the 1960s  by Chandrasekhar \citep{1964PhRvL..12..114C,1964ApJ...140..417C}, and they lead to a Sturm-Liouville problem. For numerical analyses, though, it is more convenient to use a set of coupled first order differential equations derived in \cite{1997A&A...325..217G}, written in terms of the Lagrangian displacements of the pressure and the radial coordinate. 

We start with the equations for the perfect fluid case and then generalize them for elastic systems. For perfect fluids, they are \citep{1997A&A...325..217G}

\begin{equation}\label{ecuacionparaXI}
\left(\frac{d\bar{\xi}}{dr}\right)_{\rm perf}=V(r)\bar{\xi}+W(r)\Delta p,
\end{equation}
\begin{eqnarray}\label{ecuacionparaP}
\left(\frac{d\Delta p}{dr}\right)_{\rm perf}
= X(r) \bar{\xi} + Y(r)  \Delta p, 
\end{eqnarray}
with
\begin{eqnarray}
V(r) &=& -\frac{3}{r}-\frac{dp}{dr}\frac{1}{(p+\rho)},  \\  
W(r) &=& -\frac{1}{r}\frac{1}{\Gamma p},   \\
X(r) &=& \omega^{2}e^{\lambda-\nu}(p+\rho)r + \bigg(\ \frac{dp} {dr}\bigg)^{2}\frac{r}{(p+\rho)}\nonumber\\&-&8\pi e^{\lambda}(p+\rho) pr -4\frac{dp}{dr},    \\
Y(r) &=& \frac{dp}{dr}\frac{1}{(p+\rho)}- 4\pi(p+\rho)r e^{\lambda} \label{coefecuacionparaP},
\end{eqnarray}
where $\bar \xi \equiv \Delta r/r$($\propto e^{i\omega t}$), $\Gamma$ is the adiabatic index of a star and $\Delta p(\propto e^{i\omega t})$ is the Lagrangian displacement of the pressure; for hybrid stars $\Gamma$ is in general a discontinuous distribution. The functions $V(r)$, $X(r)$ and $Y(r)$ are related to background quantities, satisfying the Tolman-Oppenheimer-Volkoff (TOV) system of equations, namely,
\begin{eqnarray}
\label{tov1}
&&\frac{dp}{dr} = - \frac{\rho m}{r^2}\bigg(1 + \frac{p}{\rho}\bigg)
	\bigg(1 + \frac{4\pi p r^3}{m}\bigg)\bigg(1 -
	\frac{2m}{r}\bigg)^{-1},
\\ \nonumber \\
\label{tov3}
&&\frac{dm}{dr} = 4 \pi r^2 \rho,
\end{eqnarray}
where $p$ is the pressure, $\rho$ is the mass-energy density and $m$ the gravitational mass at $r$. The background metric is given by the Ansatz
\begin{equation}
\label{dsz_tov}
ds^{2}=-e^{ \nu(r)} dt^{2} + e^{ \lambda(r)} dr^{2} + r^{2}(d\theta^{2}+\sin^{2}{\theta}\;d\phi^{2}),
\end{equation}
with $\nu$ given by
\begin{equation}
\label{tov2}
\frac{d\nu}{dr} = - \frac{2}{\rho + p} \frac{dp}{dr},
\end{equation}
whereas the function $\lambda(r)$ is related with  $m(r)$ by means of
\begin{equation}
e^{\lambda(r)}=\left[1-\frac{2m(r)}{r} \right]^{-1}.
\label{lambda}
\end{equation}

\subsection{Adiabatic index and reaction timescales}

It is important to note that the adiabatic index, defined in general as 
\begin{equation}
\Gamma \equiv \frac{n_b}{p} \frac{\Delta p}{\Delta n_b}\label{Gamma},
\end{equation}
where $n_b$ is the baryon number density, need not be taken as the equilibrium value, $\Gamma_{\rm{eq}}$ (in the presence of perturbations). Its actual value depends upon weak reactions in the 
NS matter \citep{1989A&A...217..137H,1995A&A...294..747G,Haensel:2002qw}. In particular, when weak reaction timescales, $\tau_{\rm{weak}}$, are much smaller than the one of perturbations, $\tau_{\rm{pert}}$, matter is in weak equilibrium at all times, and one is allowed to take $\Gamma= \Gamma_{\rm{eq}}$ \citep{1995A&A...294..747G}. However, if $\tau_{\rm{weak}}\gg \tau_{\rm{pert}}$, then one must take $\Gamma= \Gamma_{\rm{frozen}}$ \citep{1995A&A...294..747G}, where ``frozen'' stands for a situation where matter is not in full thermodynamic equilibrium, and its composition does not change with perturbations. For practical purposes, it remains in a metastable equilibrium state. Clearly, the full thermodynamic equilibrium and the frozen composition are idealized cases. Real systems would have adiabatic indices between $\Gamma_{\rm{eq}}$ and $\Gamma_{\rm{frozen}}$ \citep{Haensel:2002qw}.

Let us consider first the issue of weak reactions in the hadronic phase. Processes such as Urca and modified Urca lead to $\tau_{\rm{weak}}> \tau_{\rm{pert}}$ \citep{Haensel:2002qw}, and therefore are too slow to get 
$\Gamma=\Gamma_{\rm{eq}}$. Notwithstanding, it has been shown that in the regime of densities smaller than 
the threshold for the presence of hyperons---where the hadronic phase would usually live for first order phase transitions in hybrid stars--- $(\Gamma_{\rm{frozen}}-\Gamma_{\rm{eq}})/\Gamma_{\rm{eq}}\lesssim 15\%$ \citep{Haensel:2002qw}. The  difference obtained in \cite{Haensel:2002qw} is on the  high side.  For  the  SLy4  model of the hadronic phase, used in the present paper, $(\Gamma_{\rm{frozen}}-\Gamma_{\rm{eq}})/\Gamma_{\rm{eq}}\lesssim 5\%$ (see Fig. 3 
of \citep{2001A&A...380..151D}).

In order to estimate  $(\Gamma_{\rm{frozen}}-\Gamma_{\rm{eq}})/\Gamma_{\rm{eq}}$  for the quark core, we have used the approach  applied in \cite{2006NCimB.121.1349H} within the MIT bag model for uds quark matter.
At  $\rho\sim 10^{15}$~g~ cm$^{-3}$, the dependence of $(\Gamma_{\rm{frozen}}-\Gamma_{\rm{eq}})/\Gamma_{\rm{eq}}$ on the QCD coupling constant and $B$ was very weak, while the dependence on the quark strange mass, $m_s$, was strong ($\propto m_s^6$). As a reference, for the quark mass $m_s=200\;{\rm MeV/c^2}$,
we get $(\Gamma_{\rm{frozen}}-\Gamma_{\rm{eq}})/\Gamma_{\rm{eq}}\simeq 0.5\%$. Therefore, the equilibrium adiabatic index is a good approximation for the adiabatic index of the quark phase in the presence of perturbations.

Finally, the assumption  $\Gamma = \Gamma_{\rm{eq}}$ is expected to result in the minimum violation of the usual dynamical stability rules due to the quark phase elasticity. (We already know that the stability rules are violated when $\Gamma = \Gamma_{\rm{frozen}}$ for a liquid hadronic phase \citep{1995A&A...294..747G}.) Thus, this choice is important for back-of-the-envelope estimates which give the order of magnitude of the effect and single out relevant cases for future analysis. Due to this and the simplicity in obtaining $\Gamma_{\rm{eq}}$, we follow the above-mentioned route in this work.

\subsection{Elastic stresses}

We advance now to the case where some star phases might be elastic. We assume that this could just be the case in the presence of perturbations (unstrained backgrounds). In other words, the background phases of stars are perfect fluids and their aspects are obtained normally via the TOV system of equations. In order to investigate the dynamical stability regions in the $M(R)$ or $M(\rho_c)$ relations, we restrict our analysis to radial perturbations. In this case, shear forces manifest themselves as restoring forces due to the radial Lagrangian displacements of volume elements from their equilibrium positions. We follow the shear description given by \cite{2011PhRvD..84j3006P}, which we rewrite to the radial case (for more details on the formalism, see \cite{2019CQGra..36j5004A} and references therein). In the presence of perturbations, the stress-energy momentum tensor gains a (Eulerian perturbed) component due to shear stresses and it is given by \citep{2011PhRvD..84j3006P,2019CQGra..36j5004A}
\begin{equation}
\delta \Pi^{b}_a= -\tilde{\mu} \left({\cal P}^c_aP^{db}-\frac{1}{3}{\cal P}_a^b{\cal P}^{cd}\right)\Delta g_{cd} \label{delta_pi},
\end{equation}
where $\tilde{\mu}$ is the shear modulus, ${\cal P}_{ab}\equiv g_{ab}+u_au_b$ is the projector onto the orthogonal direction of the four-velocity $u^a$, and 
\begin{eqnarray}
\Delta g_{cd}&=& h_{cd} + \xi_{c;d}+ \xi_{d;c}\nonumber\\
&=&h_{cd} + \partial_c\xi_d + \partial_d\xi_c-2\Gamma^a_{cd}\xi_a\label{Delta_metric}.
\end{eqnarray}
In the above equation use has been made of the shortcut $\partial_c\equiv \partial/\partial x^c$ and $\Gamma^a_{cd}$ are the usual Christoffel symbols (connection coefficients) \citep{1975ctf..book.....L}. 
Note here that $\xi^a\equiv\Delta x^a=(0,\xi,0,0)$ and $\xi_b=g_{ab}\xi^a$. In addition, from Eq.~\eqref{dsz_tov},
\begin{eqnarray}
h_{ab} &\equiv& \delta g_{ab}\nonumber \\
&=& \mbox{diag}(-e^{\nu_0(r)}\delta\nu(r,t), e^{\lambda_0(r)}\delta\lambda(r,t),0,0)\label{hab},
\end{eqnarray}
where $\delta\lambda$ and $\delta\nu$ are functions to be fixed by the perturbed (first order) Einstein equations and would generalize the quantities valid for perfect fluids (see, e.g., \cite{1973grav.book.....M,1964ApJ...140..417C}). For the case of radial oscillations, it is also simple to show that from the definition of $u^a$ and $u^au_a=-1$,
\begin{equation}
u^a= e^{-\frac{\nu_0}{2}}\left[\left(1-\frac{1}{2}\delta\nu\right), \dot{\xi},0,0 \right].
\end{equation}
where we have defined the ``dot'' operation as the time derivative $(\dot{A} \equiv \partial_t A)$.

By using the Christoffel symbols for the metric \eqref{dsz_tov} and keeping terms up to first order in $\xi$, it follows that the only nonzero components of $\delta \Pi^b_a$ are the diagonal ones $\left(F'\equiv {dF}/{dr} \right)$,
\begin{equation}
\delta \Pi^r_r= -\frac{2\tilde{\mu}}{3r}[r\delta\lambda + \xi(r\lambda_0'-2)+2r\partial_r\xi]\label{delta_pi_11},
\end{equation}
and $\delta\Pi^{\theta}_{\theta}=\delta\Pi^{\phi}_{\phi}=-\delta\Pi^r_r/2$. Just for completeness,
\begin{equation}
\delta T^b_a= (\delta T^b_a)_{\rm{perf}} + \delta \Pi^b_a \label{delta_T_ab_total},
\end{equation}
where the first term is the perturbation of the energy momentum tensor of the perfect-fluid background.

\subsection{Generalized perturbation equations}

Now we proceed with the generalization of the radial perturbation equations for elastic stars. Note first that $\delta\Pi^b_t=0$ implies that the $[rt]$ component of the Einstein equations does not change with respect to a perfect fluid. Thus \citep{1973grav.book.....M,1964ApJ...140..417C},

\begin{equation}
\delta \lambda = -8\pi r e^{\lambda_0}(p_0+\rho_0)\xi=-(\lambda_0+\nu_0)'\xi \label{delta_lambda},
\end{equation}
where the subscript ``$0$'' has been used for the background quantities. In the second equality of the above equation, use has been made of the background field equations. From the $[tt]$ components of the Einstein equations, it also follows that \citep{1973grav.book.....M,1964ApJ...140..417C}
\begin{equation}
\delta\rho = -r^{-2}\partial_r[r^2(p_0+\rho_0)\xi]\label{delta_energy}.
\end{equation}
From the perturbed $[rr]$ components of the Einstein equations and Eq. (\ref{delta_lambda}), one has that 
\begin{equation}
\delta\nu' = 8\pi r e^{\lambda_0}\left[ \delta p + \delta\Pi^r_r-(p_0+\rho_0)\left(\nu_0'+\frac{1}{r}\right)\xi \right]\label{delta_nu_prime},
\end{equation}
which could also be further simplified by the background equations, but we choose not to do so here. 

One could also work with other Einstein equations for obtaining the final pulsation equation, but it turns out to be more efficient to work with $T^a_{r\; ;\,a}=0$. In the spherically symmetric case, it implies that
\begin{eqnarray}
\partial_tT^t_r + \frac{T_r^t}{2}\partial_t(\nu+\lambda) + \partial_rT^r_r + \frac{2}{r}(T^r_r-T^{\theta}_{\theta})
+ \frac{\partial_r\nu}{2}(T^r_r-T^t_t)=0. \nonumber\\  \label{div_T_spherical}
\end{eqnarray}
In the case of radial perturbations, it is easy to show that \citep{1964ApJ...140..417C,1973grav.book.....M}
\begin{equation}
T^t_r= -e^{\lambda_0-\nu_0}T^r_t=(p_0+\rho_0)\dot{\xi}e^{\lambda_0-\nu_0}\label{T01}.
\end{equation}
From Eulerian perturbations, 
\begin{equation}
\nu=\nu_0+\delta\nu,\;\;\lambda=\lambda_0+\delta\lambda,\;\;T_{a}^{b}= (T^{a}_{b})_0+\delta T^{a}_{b}\label{Pertubations_Euler}.
\end{equation}
One already has all ingredients to calculate $\delta T^a_b$. From the perfect-fluid background and Eq. \eqref{delta_T_ab_total}
\begin{eqnarray}
T^t_t &=& -\rho_0 - \delta \rho,\quad 
T_r^r= p_0+\delta p+ \delta\Pi^r_r,\nonumber\\
T_{\theta}^{\theta}&=&p_0+\delta p+ \delta\Pi^{\theta}_{\theta}\label{Pert_euler_T}.
\end{eqnarray}
(We note that under shear stresses a star becomes anisotropic, and our model could be seen as particular realization of the more general analysis of \cite{2019PhRvD..99j4072R}. Indeed, as we shall show, more compact stars could emerge in this case.) Therefore, when Eqs. \eqref{T01}, \eqref{Pertubations_Euler} and \eqref{Pert_euler_T} are replaced into Eq. \eqref{div_T_spherical}, one has $(\delta\Pi^{\theta}_{\theta}=-\delta\Pi^r_r/2)$
\begin{eqnarray}
(&p_0&+\rho_0)e^{\lambda_0-\nu_0}\ddot{\xi} + \partial_r(\delta p+ \delta\Pi^r_r) + \frac{3}{r}\delta\Pi^r_r \nonumber\\
&+& \frac{\delta\nu'}{2}(p_0+\rho_0)+\frac{\nu'_0}{2}(\delta p+ \delta\rho + \delta \Pi^r_r)=0\label{eq_pert_final}.
\end{eqnarray}
Note that the above equation is the general equation for radial perturbations in elastic stars once $\delta p$ is found, given that all other terms are already known. Actually, for adiabatic processes, $\delta p$ can be easily obtained. Indeed, from Eq.\eqref{Gamma}, one learns that
\begin{equation}
\delta p \equiv \Delta p - p_0'\xi= \Gamma p_0 \frac{\Delta n_b}{n_b}- p_0'\xi \label{delta_p_interm.}.
\end{equation}
Given the perturbations, one can always find $\Delta n_b/n_b$ geometrically by means of \citep{2011PhRvD..84j3006P,2019CQGra..36j5004A}
\begin{equation}
\Delta n_b = -\frac{n_b}{2}{\cal P}^{ab}\Delta g_{ab}\label{Delta_nb}.
\end{equation}
In the spherically symmetric case,
\begin{eqnarray}
-\frac{\Delta n_b}{n_b}&=& \frac{\delta \lambda}{2} + \frac{2}{r} \xi + \frac{\lambda_0'}{2}\xi + \partial_r \xi \nonumber\\
&=& \frac{1}{r^2}e^{-\frac{\lambda_0}{2}} \partial_r\left( e^{\frac{\lambda_0}{2}}r^2\xi\right)+\frac{\delta \lambda}{2} \nonumber\\
&=&\frac{1}{r^2}e^{\frac{\nu_0}{2}} \partial_r\left( e^{\frac{-\nu_0}{2}}r^2\xi\right)
\label{Delta_nb_final},
\end{eqnarray}
where, in the last equality of the above equation, we have used Eq.~\eqref{delta_lambda}. Formally, Eq.~\eqref{Delta_nb_final} is the same as its perfect-fluid counterpart.

As in the case of perfect fluids, for elastic stars it is also numerically convenient to solve equations for $\xi$ and $\Delta p$ as if they were independent variables. This can be easily found with the help of Eqs. \eqref{eq_pert_final}, \eqref{delta_p_interm.} and \eqref{Delta_nb_final}. From Eqs. \eqref{delta_p_interm.} and \eqref{Delta_nb_final}, when the variable $\bar{\xi}\equiv \xi/r$ is inserted, it follows that [($\bar{\xi}, \Delta p) \propto e^{i\omega t}$ is assumed from now on]
\begin{equation}
\frac{d\bar{\xi}}{dr}=-\frac{1}{r} \left(\frac{\Delta p}{p_0\Gamma} + 3\bar{\xi} \right)-\frac{p_0'\bar{\xi}}{p_0+\rho_0}\label{xi_bar_eq},
\end{equation}
which is exactly Eq. \eqref{ecuacionparaXI}. For obtaining the above equation, use has been made of Eq. \eqref{tov2}. The second equation for the coupled system of equations involving $\bar{\xi}$ and $\Delta p$ can be easily found from Eq. \eqref{eq_pert_final} when the first equality of Eqs.~\eqref{delta_p_interm.} and \eqref{delta_nu_prime} are taken into account. 
It can be schematically written as
\begin{eqnarray}
\frac{d\Delta p}{dr}&=& \left(\frac{d\Delta p}{dr} \right)_{\rm{perf}} -\ \frac{d\delta\Pi^r_r}{dr}\nonumber\\ &-& \left[\frac{3}{r} + 4\pi re^{\lambda_0}(p_0+\rho_0) +\frac{1}{2}\nu_0'\right]\delta\Pi^r_r.\label{Delta_p_perf.fluid}
\end{eqnarray}
Naturally, in addition to the perfect-fluid part [see Eq. \eqref{ecuacionparaP}], several other terms also appear due to the elasticity of the star. In order to see their influence on $d\Delta p/dr$, we rewrite $\delta \Pi^r_r$ based on Eq.~\eqref{xi_bar_eq}. With the use of Eq.~\eqref{delta_lambda}, Eq.~\eqref{delta_pi_11} simplifies to 
\begin{equation}
\delta \Pi^r_r= \frac{4}{3}\tilde{\mu}\left(\frac{\Delta p}{p_0\Gamma} +3\bar{\xi} \right). \label{Delta_p11_final}
\end{equation}

\subsection{Boundary and matching conditions to elastic hybrid stars}
We now set out the boundary and matching conditions to the problem of radial perturbations in elastic hybrid stars. It turns out not all of them are the same as in the perfect-fluid case \citep{2018ApJ...860...12P}. Let us explicitly show this. Equation~\eqref{xi_bar_eq} is only well defined at the origin if 
\begin{equation}
\Delta p (0)= -3\bar{\xi}(0)p_0(0)\Gamma(0) \label{cond_Delta_p},
\end{equation}
which is always meaningful when $\bar{\xi}(0)$ is finite. 
From Eq.~\eqref{Delta_p_perf.fluid}, we note that one has also to impose that $\delta \Pi^r_r(0)=0$. However, this is automatically satisfied through Eq. (\ref{cond_Delta_p}). At the surface of the star ($r=R$), one should impose that
\begin{equation}
\Delta p(R)=0\label{cond_Delta_p_surface},
\end{equation}
with the obvious condition that $\bar{\xi}(R)$ is finite. 

In hybrid stars, one should also provide matching conditions to $\bar{\xi}$ and $\Delta p$ at the phase-splitting surfaces.

Here we choose to work with nucleation processes around these interfaces. Essentially, they could be classified into two asymptotic regimes: when the timescales of nucleation ($\tau_{\rm{nucl}}$) are either much larger or much smaller than the ones of perturbations ($\tau_{\rm{pert}}$). When $\tau_{\rm{nucl}}\ll \tau_{\rm{pert}}$, volume elements are instantaneously converted from one phase to the other and the phase conversion is said to be ``rapid''. For the case $\tau_{\rm{nucl}}\gg \tau_{\rm{pert}}$, nucleation processes occur at a very low speed and for all purposes volume elements just squash or stretch when perturbed. For this reason we call them ``slow phase conversions''.

Just for definiteness, assume that a phase-splitting surface in a hybrid star is at $r=R_{\rm{pt}}$ (``pt'' stands for ``phase transition'') and jumps of physical quantities are with respect to it. It has been shown \citep{2018ApJ...860...12P,1989A&A...217..137H} that for slow (s) phase conversion processes, the extra-matching condition to be taken for $\bar{\xi}$ is 
\begin{equation}
[\bar{\xi}_s]^+_-\equiv \lim_{q\rightarrow 0^+}\left[\bar{\xi}_s\right]^{R_{\rm pt} + q}_{R_{\rm pt}-q} =0\label{xi_slow},
\end{equation}
while for rapid (r) phase transitions,
\begin{equation}
[\bar{\xi}_r]^+_-= \left[\frac{\Delta p}{r p_0'} \right]^+_-\label{xi_rapid}.
\end{equation}

Naturally, Eq. \eqref{xi_slow}  is not only related to slow nucleation processes. It is a standard matching condition to be taken into account when there is no mass fluxes from one phase to the other. Slow conversions happen to be a special case where these conditions are fulfilled. 

We turn now to the jump condition for $\Delta p$ at $R_{\rm pt}$. This is obtained as a byproduct of the perturbation equations, and it could be more easily obtained from Eq. \eqref{Delta_p_perf.fluid}. When it is promoted to a distribution, the Dirac delta terms in Eq. \eqref{Delta_p_perf.fluid} imply that $[\Delta p - (\Delta p)_{\rm{perf}} + \delta\Pi^r_r]^+_-=0$. Since for perfect fluids $[(\Delta p)_{\rm{perf}}]^+_-=0$ \citep{2018ApJ...860...12P,2020arXiv200301259T}, it trivially follows that
\begin{equation}
[\Delta p + \delta\Pi^r_r]^+_-=0\label{jump_Delta_p}, 
\end{equation}
which is a consequence of the continuity of the radial traction at $R_{\rm pt}$ \citep{2020arXiv200310781P}. Hence, when some of the parts of the star are elastic, the Lagrangian displacement of the pressure at the background phase-splitting surface is discontinuous. From Eq. \eqref{Delta_p11_final}, Eq. \eqref{jump_Delta_p} is actually an algebraic equation for $[\Delta p]^+_-$, which could be easily solved for both slow [Eq.~\eqref{xi_slow}] and rapid [Eq. \eqref{xi_rapid}] phase conversions. A non-null jump of $\Delta p$ on top of Eqs. \eqref{xi_slow} and \eqref{xi_rapid} might also have important implications for the dynamical stability of stars with elastic parts. We recall that it is already known that when $\Delta p$ is continuous, Eq. \eqref{xi_slow} results in $\omega_s=0$ for $\partial M_0/\partial \rho_c<0$ and Eq. \eqref{xi_rapid} leads to the presence of the reaction mode \citep{1989A&A...217..137H,2018ApJ...860...12P} (an extra-mode which does not have an one-phase counterpart), though for rapid phase conversions $\partial M/\partial \rho_c=0$ when $\omega_r=0$. At least one would expect quantitative changes of these results in the presence of elasticity.

Put in the above way, the problem of radial perturbations in stars with elastic parts and phase transitions is a Sturm-Liouville problem. Thus, the set of characteristic eigenfrequencies $\omega$ is discrete and hierarchic ($\omega^2_0<\omega^2_1<\omega^2_2....$, where $\omega^2_i$ is the $i$th-mode). For dynamical stability purposes, it suffices to investigate the eigenfrequencies of the fundamental mode. 

The appropriate quark-hadron matching condition critically depends on the physical processes and scenarios one wants to investigate.  When a hadronic fluid element  in the neighbourhood of the sharp interface is compressed beyond the transition pressure, a rapid conversion to LOFF quark matter is expected to be strongly suppressed at low enough temperatures.  To understand this, let us focus first on unpaired quark matter.  Typical timescales for nucleation driven by quantum fluctuations are much larger than the age of the Universe for temperatures below $10 \, \mathrm{MeV}$  \citep{2016EPJA...52...58B}. If thermal  nucleation is considered, the timescale is also larger than the age of the Universe for $T<5 \, \mathrm{MeV}$ but decreases to $\sim 1$ s for $T \approx 10 \mathrm{MeV}$ \citep{2016EPJA...52...58B}. These numbers are significantly larger than the typical period of radial oscillations strongly suggesting that the assumption of slow phase conversions is appropriate for cold enough matter.  In the case of quark matter in a LOFF crystalline phase, this conclusion is expected to remain valid because an additional time (with respect to unpaired matter) would be needed to form the specific three-dimensional lattice structure needed for LOFF formation \citep{2006PhRvD..74i4019R}. As stressed previously, nucleation timescales can decrease significantly for hot NSs. It has been suggested that they may be smaller than typical perturbation timescales for temperatures around 20 MeV \citep{2020arXiv200301259T}. In this case, conversions around a phase-splitting surface should be rapid. Examples of stars fulfilling these conditions would be those in the aftermath of binary NS mergers. However, a natural caveat here would be the need to work with hot equations of state. 

Given that we are mainly interested in cold stars (both due to the possibility of probing the crystalline color-superconducting phase and the fact that such stars are the main targets of missions such as NICER and eXTP), we mostly focus our analysis on slow conversions. Nonetheless, for completeness, we also comment on rapid conversions.

\section{Results}
\label{results}

\begin{figure}[htbp]
\centering
\includegraphics[width=1.05\columnwidth,clip]{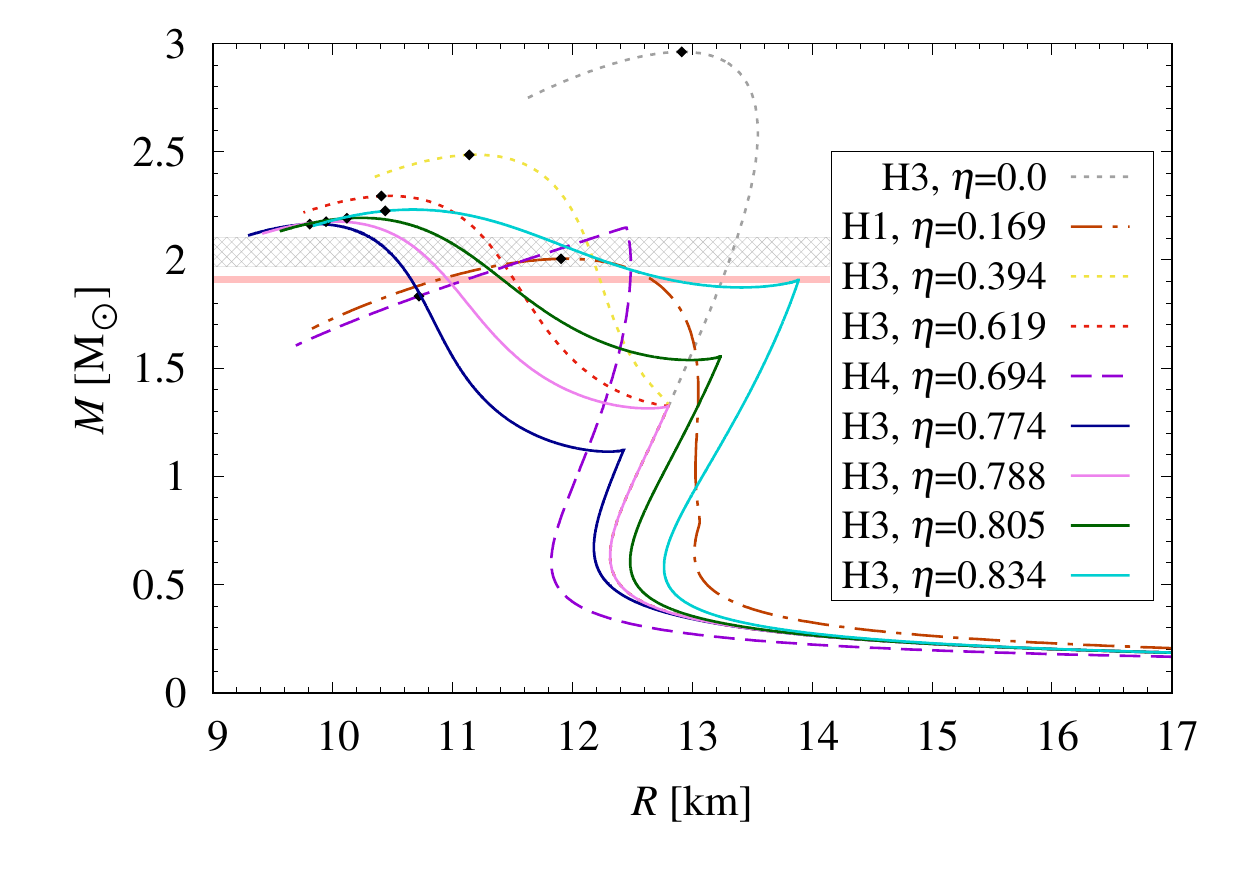}
\caption{Mass-radius $M(R)$ relations for the hybrid star models of Table \ref{table_models}. We mostly focus on hybrid models with the SLy4 EOS for different density jumps (H3 models), but also consider the chiral effective theory (H1 and H4 models). From bottom to top of the $M(R)$ thick curves (from gray to red) we have chosen $n_0=\mathrm{(0.235,0.21,0.185,0.16) \ fm^{-3}}$. The dashed lines have the same $n_0$ as the thick yellow line ($n_0=\mathrm{0.21 \ fm^{-3}}$), but different density jumps. For all H3 models, we take $n_1=0.335$ fm$^{-3}$ ($\sim 2.1 \rho_{\rm sat}$, the density marking the end of the hadronic phase). The upper horizontal band corresponds to the observed mass of the pulsar PSR J0348+0432, while the lower one relates to PSR J1614-2230 \citep{arzoumanian2018,2013Sci...340..448A}. The diamonds on each EOS mark the terminal masses ($\omega^2=0$) for perfect-fluid hybrid stars under slow conversions. Stable stars are on the right of the markers. For elastic hybrid stars, $\omega^2=0$ is always more to the left of the diamonds and their precise locations depend on the shear modulus of the quark phase. 
}

\label{fig:MR}
\end{figure}

In the forthcoming analysis we do not make a systematic study of the space of parameters of the quark phase, but just choose some representative parameters. We choose $(B,a_4, a_2)$ in such a way to respect the constraints already measured for NSs such as tidal deformabilities and maximum masses, as well as assume that the hadronic phase generally starts at densities larger than the nuclear saturation density ($\rho_{\rm sat} = 2.7\times 10^{14}$ gcm$^{-3}$). 
For the hadronic phase, we work either with chiral models or the SLy4 + polytropic equation of state, as explained in Sec. \ref{models_hybrid_stars}. Table \ref{table_models} summarizes the main aspects of the models used in this work and Fig. \ref{fig:MR} shows some associated $M-R$ relations for hybrid stars. Just to check the consistency of numerical calculations, we also make use of the  (book-keeping) polytropic EOS $p=K\rho^2$, with $K=100$ km$^2$ for the hadronic phase. Details about the connection (through the Maxwell construction) of this EOS with the MIT bag-like model are given in \cite{2020arXiv200310781P}. We compute the tidal deformability ($\Lambda$) based on perfect fluids (we follow the formalism of \cite{2008ApJ...677.1216H,2009PhRvD..80h4035D,2010PhRvD..81l3016H}), even though we are assessing elastic stars. The reason is because an elastic star should have a smaller Love number than a perfect-fluid star \citep{2011PhRvD..84j3006P,2020PhRvD.101j3025G,2020arXiv200310781P} and if the latter already satisfies the GW constraints, so will the former. In what follows, we focus more on the case of slow conversions given that they would be expected for cold stars. We use the phenomenological shear modulus given by Eq. \eqref{mu-quark-p} in order to reflect the uncertainties in the calculations of \cite{2007PhRvD..76g4026M}.

\begin{center}
\begin{table}
[htbp] 
\begin{ruledtabular}
{\begin{tabular}{@{}c|cccc@{}} 
Hybrid& Quark Model & $\eta$ & Hadronic \\ 
Model&$\left(\frac{B^{\frac{1}{4}}}{\mbox{MeV}},a_4, \frac{a_2^{\frac{1}{2}}}{\mbox{MeV}}\right)$ & $\left(\frac{\rho_q}{\rho_h}-1\right)$& Model \\ \hline
H1&$(142.39,0.55,80)$ & 0.169 & Chiral+Polytr. (stiff)\\ 
H2&$(137,0.4,100)$ & free & Polytr. (n=1)\\
H3& bag with $c_s^2=1$ &free & SLy4+Polytr. ($\gamma=4.5$)\\ 
H4& $(163.30,0.7,150)$ &0.694 & Chiral+Polytr. (intermed.)\\ 
\end{tabular}}
\end{ruledtabular}
\caption{Main aspects of the hybrid star models used in our work. For the chiral model, stiff and intermediate EOSs have been taken into account \citep{Hebeler:2013nza}. Our notation is such that $\rho_q$ is the density at the top of the quark phase and $\rho_h$ is the density at the bottom of the hadronic phase. Therefore, $\eta$ is directly related to the density jump at the quark-hadron phase-splitting surface ($[\rho]^+_-\equiv \rho_h-\rho_q$).}
\label{table_models}
\end{table}
\end{center}

For cold, catalyzed stars with no phase transitions (one-phase stars), powerful theorems deduced in \cite{Wheeler1965} state that for dynamically stable stars, $\partial M/\partial \rho_c\geq 0$. Here $M$ is the background (gravitational) mass of a star with a given EOS and a central density $\rho_c$. It is already known that one-phase stars under slow weak reactions (slightly) violate this condition \citep{1995A&A...294..747G}. 
However, appreciable violations might happen for hybrid stars under slow phase conversions \citep{2018ApJ...860...12P},  another reason why we choose to investigate them. 

When it comes to the stability of stars, the terminal configurations (for which $\omega^2 = 0$) are of main interest. We have employed the numerical method of \citet{Mariani}. In short, we set $\omega ^2 = 0$ in the radial oscillation equations and solve them together with the TOV equations. The Lagrangian pressure $\Delta p$ must be zero at the stellar surface, and we can find the configuration satisfying $\omega^2 = 0$ and $\Delta p (R) = 0$ simultaneously by varying the central density/pressure. Due to the frequency ordering, if we start from the hadronic branch and increase the central pressure, the first mode to be found is necessarily the fundamental one. 
A crucial point is to fulfill the appropriate matching conditions at the quark-hadron interface. In order to do that, we integrate the equations phase by phase, starting from the quark core. We also split the integrations in the hadronic phase in order to take into account its elastic and liquid ocean regions (see section \ref{hadronic_phase} for the transition densities).
However, the precise densities limiting the elastic hadronic phase are actually not very important because its effects on the stability of the stars turn out to be very small.

Table \ref{ta_results} summarizes some stability aspects of hybrid stars with elastic quark phases for $\alpha=1$ and $\Delta=25$ MeV ($\tilde{\mu}_{q, 400}^{(p)}\sim $ 15 MeV fm$^{-3}$). (Table \ref{table_pt_parameters} gives the phase transition parameters for the H1, H3 and H4 EOSs.) The numbers are obtained via the numerical solution of the perturbation equations from Sec. \ref{model-radial-perturbations-solid-stars} with different matching conditions (slow and rapid phase conversions) as described above. One sees that relative radius changes for the zero eigenfrequency for both conversions are $0.1-2$\%. (Slow conversions lead to relative radius changes smaller than rapid ones.) Such differences are with respect to the radii of null eigenfrequencies of perfect-fluid stars. Observationally speaking, this would translate in the following way. Due to elasticity and slow conversions, a given observed mass between $M^{\rm elas}_{\omega=0}$ and $M^{\rm perf}_{\omega=0}$ could now be on either side of the maximum of the $M-R$ relation. Thus, the above changes roughly mean that the expected relative radius dispersion {\it exclusively} due to a quark phase elasticity for stars with masses in the above mass range would be up to $0.2-4$\%.

Larger radius differences are also possible for larger values of $\alpha$. Table \ref{table_radius_change_maximum} shows the associated changes for $\alpha=4$ and slow conversions. 
Stiff EOSs suggest that the maximum increase of relative radius differences solely due to elastic quark aspects will surpass the $5\%$ threshold 
for  $\tilde{\mu}_{q,400}^{(p)} \sim 50$ MeV fm$^{-3}$. Soft EOSs---such as some representatives of the H3 model---suggest that the largest radius difference increases due to the elasticity of the quark phase are around $2-3\%$. Possible ways to connect the above-mentioned radius changes with observations and associated difficulties are discussed in the next section.

Tables \ref{ta_results} and \ref{table_radius_change_maximum} also suggest a nontrivial consequence of slow conversions for EOS models similar to the H4 model: masses that result in $\omega^2_{\rm s}=0$ may also be very different from the maximum one. 
\footnote{It should be mentioned that the physics around the maximum mass for the H4 model is different from the other models we have considered. It resembles more the situation around the phase transition masses (small quark cores) for the H1, H2 and H3 models in the case of large energy jumps. Slow conversions render non-negligible regions of their $M-R$ relations meta-stable when $\partial M/\partial \rho_c<0$, similarly to what happens for the H4 model. This is the reason why for the H4 model the terminal mass for a perfect-fluid star is around 15$\%$ smaller than its maximum mass, and radius differences for a mass within this extended branch could be up to $2$\,km.} However, for the H4 model in particular, the terminal radius and mass of elastic NSs only change by approximately $0.5\%$ with respect to their perfect-fluid counterparts. (When stiff EOSs for the chiral model are taken into account, maximum relative changes for the terminal mass and radius of elastic hybrid stars are around $1.5\%$.) We discuss more about the above class of EOSs in the next section.

One also notes from Tabs. \ref{ta_results} and \ref{table_radius_change_maximum} that the relative radius change depends on the density jump at the quark-hadron phase transition. However, for slow conversions, the relationship is not trivial. They depend, for instance, on the stiffness of the EOS, the values the chemical potential take in the quark phase, the relative size of the crust (when compared to the core), and where in the $M-R$ diagram the terminal radius is reached for a perfect-fluid star. \footnote{For rapid conversions, there is a simple trend for the relative radius change (due to elasticity) and the quark-hadron density jump---mostly owing to the fact that for perfect fluids the terminal radius is always related to the maximum mass. Around such a mass, a larger energy jump would imply a smaller hadronic phase and hence stars ``on the way'' of being one-phased; in this one-phased case, it is known that the usual dynamical stability rule holds \citep{2004CQGra..21.1559K}. Thus, relative radius changes in hybrid stars should decrease when the density jump increases.} In addition, it has a dependency on the EOS of the quark phase. For the H3 models, changes are small even for stiff EOSs, while for the H2 model, the radius differences are more pronounced for stiffer EOSs.

\begin{figure}[htbp]
\centering
\includegraphics[width=1.05\columnwidth,clip]{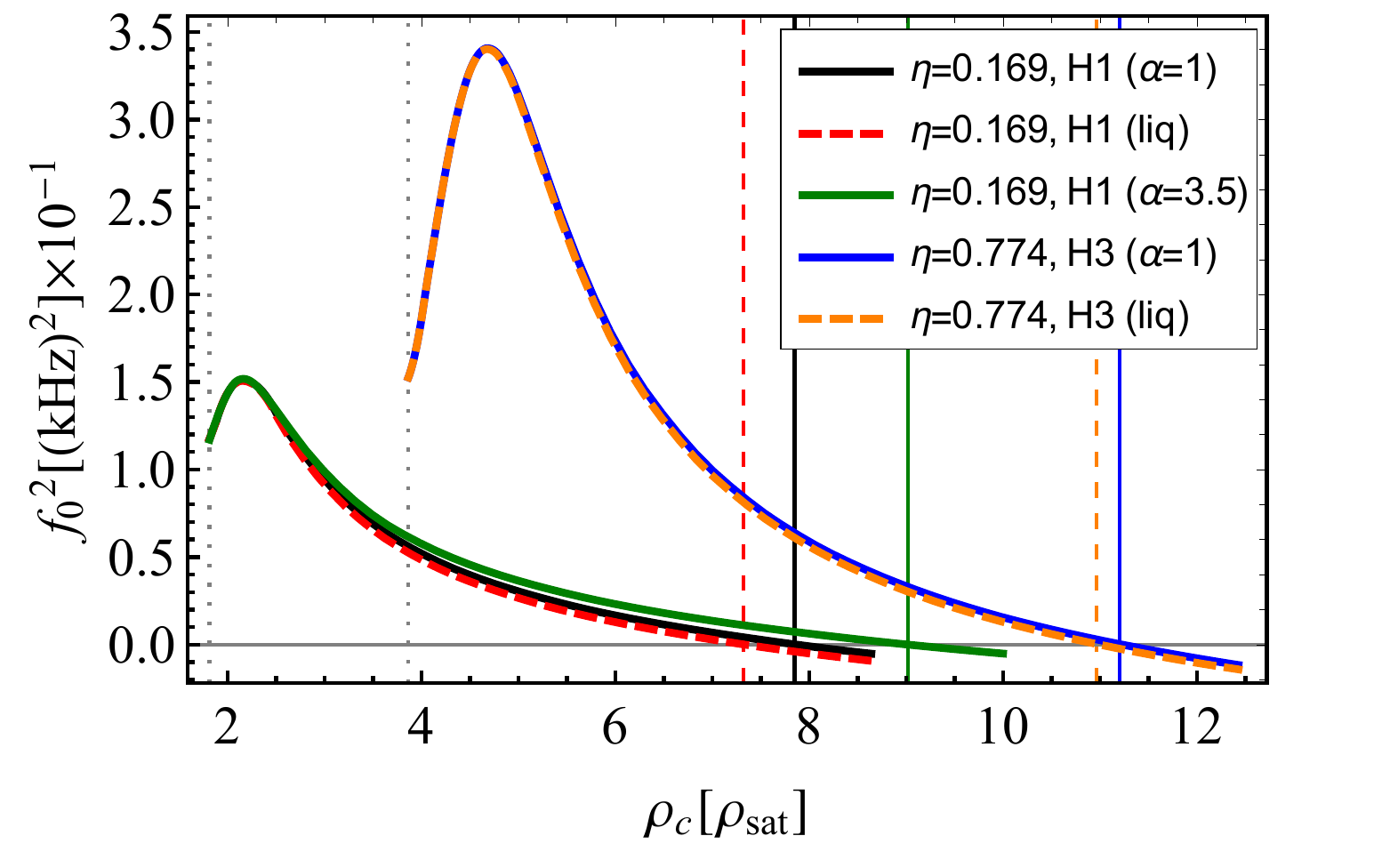}
\caption{Hybrid star fundamental frequencies ($\Delta = 25$ MeV) for the H1 and H3 models. The frequencies are associated with slow conversions, which have real values in the vicinities of strong phase transitions (\cite{2019A&A...622A.174S} and references therein), differently from rapid conversions. The non-dotted vertical lines mark the densities where the eigenfrequencies of hybrid star sequences are null (we respect the style and color of each curve for clarity). They are all larger than the densities respectively associated with their maximum masses ($\rho_c= 7.142$ $\rho_{\rm sat}$ for the H1 EOS and $\rho_c= 10.584$ $\rho_{\rm sat}$ for the H3 EOS).
At face value, the maximum shear modulus already leads to non-negligible changes to the eigenfrequencies at smaller densities than the ones for their maximum masses. Indeed, for the H1(H3) model with $\alpha=1$, relative frequency differences of 10$\%$ take place at $\rho_c= 5.78$ $\rho_{\rm sat}$($\rho_c= 10.17$ $\rho_{\rm sat}$), which corresponds to $M= 1.99\,M_{\odot}$($M= 2.166\,M_{\odot}$). When $\alpha=3.5$, instead, we have that for the H1 model (stiff) $10\%$ differences happen at $\rho_c= 4.41$ $\rho_{\rm sat}$ ($M= 1.92\,M_{\odot}$). In addition, $\omega_{\rm s}^2$ is zero at $\rho_c=9.02$ $\rho_{\rm sat}$ ($M=1.99\,M_{\odot}$, $R=11.60$ km). In this case, when compared to their perfect-fluid star counterpart (see Tab.~\ref{ta_results}), the largest increase of relative radius differences would be around 5.2~$\%$.}
\label{freq_radius_SLy}
\end{figure}

For completeness, we plot in Fig. \ref{freq_radius_SLy} 
the square of the fundamental eigenfrequencies of the H3 and H1 models ($\Delta=25$ MeV) for slow conversions with and without taking into account the elasticity of the quark core.
For $\alpha=1$, one sees that the imprint of elasticity on the eigenfrequencies is mostly relevant for large masses: relative changes would be larger than $10\%$ ($\sim 20\%$ for the frequency squared) for NS masses above $2.0$ $M_{\odot}$ (for stiff EOSs). Naturally, if the value of the shear modulus is larger (larger $\alpha$), significant differences between the eigenfrequencies of a liquid and an elastic quark phase take place at smaller masses. For instance, taking $\alpha =3.5$ and the H1 model, one finds that relative frequency changes above $10\%$ occur for masses larger than $1.9\, M_{\odot}$. We note that the masses (central densities) where $\omega=0$ are always larger than the ones where relative frequency changes become relevant. Thus, if direct or indirect observations of NS frequencies are possible, quark matter aspects for a given EOS could also be constrained with them. We return to this issue in the next section too.

\begin{widetext}
\begin{center}
\begin{table}[ht] 
\caption{Hybrid star models (see Table \ref{table_models}) and some of their properties for $\omega^2=0$ (limit of unstable branch), $\Delta= 25$ MeV and $\alpha=1$ [$\tilde{\mu}_{q,400}^{(p)}\sim $ 15 MeV fm$^{-3}$; see Eq. \eqref{mu-quark-p}] in the case of slow and rapid phase conversions. Here, ``elas'' stands for elastic, ``perf'' for perfect fluid, ``s'' for slow and ``r'' for rapid. In addition, $\Lambda_{1.4M_{\odot}}^{\rm{perf}}$ is the tidal deformability of a $1.4 M_{\odot}$ perfect-fluid star and $\rho_{\rm{c}}$ is the central density of an NS. All parameters have been chosen so that their maximum masses are above $2M_{\odot}$ and $\Lambda_{1.4M_{\odot}}\lesssim 660$, in agreement with gravitational and electromagnetic wave observations from NSs. We note that maximum masses in the $M-\rho_c$ relation coincide with null frequencies for rapid phase conversions in the perfect-fluid case.}
\begin{ruledtabular}
{\begin{tabular}{@{}c|c|ccc|ccc|ccc|ccc|cc@{}} 
Hybrid & $\eta$ &  $\rho_{\rm{c}, \omega=0}^{\rm{perf,r}}$ & $M_{\omega=0}^{\rm{perf,r}}$ & $R_{\omega=0}^{\rm{perf,r}}$ & $\rho_{\rm{c},\omega=0}^{\rm{perf,s}}$ & $M_{\omega=0}^{\rm{perf,s}}$ & $R_{\omega=0}^{\rm{perf,s}}$ &  $\rho_{\rm{c},\omega=0}^{\rm{elas,r}}$ & $M_{\omega=0}^{\rm{elas,r}}$ & $R_{\omega=0}^{\rm{elas,r}}$& $\rho_{\rm{c},\omega=0}^{\rm{elas,s}}$ & $M_{\omega=0}^{\rm{elas,s}}$ & $R_{\omega=0}^{\rm{elas,s}}$  & $\Lambda_{1.4M_{\odot}}^{\rm{perf}}$  \\ 
Model & &  $(\rho_{\rm{sat}})$ & $(M_{\odot})$ & $(\rm{km})$ & $(\rho_{\rm{sat}})$ & $(M_{\odot})$ & $(\rm{km})$ & $(\rho_{\rm{sat}})$ & $(M_{\odot})$ & $(\rm{km})$ & $(\rho_{\rm{sat}})$ & $(M_{\odot})$ & $(\rm{km})$& \\ \hline
H1 & 0.169 & 7.142 & 2.006 & 11.945 & 7.321 & 2.005 & 11.907 & 8.237 & 2.000 & 11.724 & 7.850 & 2.003 & 11.799 & 604.0  \\
\hline
H2 & 0.0 &  6.781 & 2.058 & 11.942 & 6.801 & 2.058  & 11.938  & 8.0148 & 2.050 & 11.696 & 7.591 & 2.054 & 11.776  & 658.1  \\
\hline
\textquotedbl &0.55 & 1.4417 & 0.278 & 11.940 & - & - & - & 1.4413 & 0.278 & 11.946 & - & - & - & 624.1  \\
\textquotedbl & \textquotedbl&  6.824 & 2.054 & 11.836 & 6.845 & 2.054  & 11.831  & 7.958 & 2.048 & 11.614 & 7.810 & 2.049 & 11.641  & \textquotedbl\\
\hline
\textquotedbl & 1.0 & 1.4359 & 0.204 & 11.537 & - & - & - & 1.4339 & 0.204 & 11.591 & - & - & - & 607.4  \\
\textquotedbl & \textquotedbl & 6.827 & 2.052 & 11.750 & 6.866 & 2.052 & 11.742 & 7.969 & 2.046 & 11.531 & 7.878 & 2.047 & 11.548 & \textquotedbl  \\
\hline
\textquotedbl & 1.5 & 1.4273 & 0.150 & 11.080 & - & - & - & 1.4254 & 0.150 & 11.163 & - & - & - & 599.4  \\
\textquotedbl & \textquotedbl & 6.843 & 2.051 & 11.676 & 6.868 & 2.051 & 11.671 & 7.964 & 2.045 & 11.465 & 7.906 & 2.045 & 11.475 & \textquotedbl  \\
\hline
H3 & 0.0 & 5.704 & 2.962 & 12.913 & 5.704 & 2.962 & 12.913 & 6.061 & 2.960 & 12.825 & 5.857 & 2.961 & 12.875 & 640.0  \\
\hline
\textquotedbl & 0.394 & 8.016 & 2.486 & 11.196 & 8.339 & 2.486 & 11.140 & 8.373 & 2.486 & 11.134 & 8.537 & 2.485 & 11.107 & 568.8  \\
\hline
\textquotedbl & 0.619 & 9.377 & 2.296 & 10.475 & 9.842 & 2.295 & 10.408 & 9.735 & 2.296 & 10.423 & 10.059 & 2.294 & 10.378 & 444.3  \\
\hline
\textquotedbl & 0.774 & 3.9240 & 1.114 & 12.300 & - & - & - & 3.9235 & 1.114 & 12.300 & - & - & -  & 195.5  \\
\textquotedbl & \textquotedbl & 10.584 & 2.166 & 9.855 & 10.967 & 2.166 & 9.810 & 10.927 & 2.166 & 9.814 & 11.201 & 2.165 & 9.783  & \textquotedbl  \\
\hline
\textquotedbl & 0.788 & 3.9991  & 1.314 & 12.629 & - & - & - & 3.9981 & 1.314 & 12.632 & - & - & -  & 329.1  \\
\textquotedbl & \textquotedbl & 10.413 & 2.178 & 10.016 & 10.969 & 2.177 & 9.946 & 10.771 & 2.177 & 9.970 & 11.196 & 2.176 & 9.918  & \textquotedbl  \\
\hline
\textquotedbl & 0.805 & 4.1077 & 1.537 & 12.984 & - & - & - & 4.1055 & 1.537 & 12.989 & - & - & -  & 123.8  \\
\textquotedbl & \textquotedbl & 10.160 & 2.195 & 10.232 & 10.976 & 2.193 & 10.120 & 10.535 & 2.194 & 10.179 & 11.196 & 2.192 & 10.091  & \textquotedbl  \\
\hline
\textquotedbl & 0.834 & 4.3570 & 1.873 & 13.409 & - & - & - & 4.3500 & 1.873 & 13.423 & - & - & -  & 132.6  \\
\textquotedbl & \textquotedbl & 9.578 & 2.233 & 10.671 & 10.995 & 2.227 & 10.440 & 9.982 & 2.232 & 10.600 & 11.204 & 2.225 & 10.410  & \textquotedbl  \\
\hline
H4 & 0.694 & 6.4462 & 2.150 & 12.455 & 17.487 & 1.833 & 10.722 &  6.4464 & 2.150 & 12.453 & 17.651  & 1.830 & 10.707 & 472.9  \\ 
\end{tabular} \label{ta_results}}
\end{ruledtabular}
\end{table}
\end{center}
\end{widetext}

\begin{center}
\begin{table}
[htbp] 
\begin{ruledtabular}
{\begin{tabular}{@{}c|cccc@{}} 
Hybrid & $\eta$ & $p_{\rm pt}$ & $\rho_{\rm pt}$ &   $M_{\rm pt}$ \\ 
Model &  &   (MeV fm$^{-3}$) & $(\rho_{\rm sat})$ & $(M_{\odot})$ \\ \hline
H1 & 0.169 & 15.37 & 1.81 & 0.77 \\
H3 & 0 & 41.85  & 2.19  & 1.33 \\ 
\textquotedbl & 0.394 & \textquotedbl & 3.05 & \textquotedbl \\
\textquotedbl & 0.619 & \textquotedbl & 3.54 & \textquotedbl \\ 
\textquotedbl & 0.788 & \textquotedbl & 3.91 & \textquotedbl \\
\textquotedbl & 0.774 & 35.03 & 3.86 & 1.12 \\
\textquotedbl & 0.805 & 50.15 & 3.97 & 1.56 \\
\textquotedbl & 0.834 & 64.73 & 4.08 & 1.91 \\
H4 & 0.694& 168.8 & 6.41 &  2.15 \\ 
\end{tabular}}
\end{ruledtabular}
\caption{Phase transition (``pt'') parameters of the H1, H3 and H4 models featuring in Table \ref{ta_results}. Here, $\rho_{\rm pt}$ is connected with $p_{\rm pt}$ via the quark equation of state, and it coincides with $\rho_q$ defined in Table \ref{table_models}. The quantity $M_{\rm pt}$ is the gravitational mass marking the appearance of the quark phase (cusps in the $M-R$ relations). Just for completeness, we note that $M_{\rm pt}$ for the H4 model is slightly smaller than its maximum mass.}
\label{table_pt_parameters}
\end{table}
\end{center}

\begin{center}
\begin{table}
[htbp] 
\begin{ruledtabular}
{\begin{tabular}{@{}c|cccc@{}} 
Hybrid & $\eta$ & $\rho_{\rm c}$ & $R^{\rm elas,s}_{\omega=0}$ & $1-\frac{R^{\rm elas,s}_{\omega=0}}{R^{\rm perf,s}_{\omega=0}}$ \\ 
Model &  & $(\rho_{\rm sat})$ & (km) & ($\%$) \\ \hline
H1 & 0.169 & 9.171 & 11.560 & 2.92 \\
H2 & 0 & 9.530 & 11.455 & 4.28 \\
\textquotedbl & 0.55 & 10.345 & 11.247 & 5.15 \\ 
H3 & 0 & 6.297 & 12.770 & 1.11 \\ 
\textquotedbl & 0.394 & 9.125 & 11.015 & 1.13 \\ 
\textquotedbl & 0.774 & 11.904 & 9.707 & 1.05 \\
\textquotedbl & 0.788 & 11.879 & 9.841 & 1.06 \\
\textquotedbl & 0.805 & 11.853 & 10.012 & 1.07 \\
\textquotedbl & 0.834 & 11.902 & 10.319 & 1.16 \\
H4 & 0.694 & 18.056 & 10.669 & 0.49 \\ 
\end{tabular}}
\end{ruledtabular}
\caption{Relative changes of the terminal radius (for which $\omega^2=0$) for hybrid star with elastic quark cores ($\Delta = 25$ MeV and $\alpha=4$: $\tilde{\mu}_{q,400}^{(p)}\sim 60 $ MeV fm$^{-3}$) in the case of slow conversions. The maximum relative radius change for a given mass range exclusively due to quark elasticity would be approximately twice as large for all the models. The largest differences relate to stiff EOSs.}
\label{table_radius_change_maximum}
\end{table}
\end{center}

\section{Discussion and conclusions}
\label{discussion}

Elastic aspects to hybrid stars make them distinctly different from their perfect-fluid counterparts because they have dissimilar restoring forces upon perturbations. This naturally influences the dynamics of perturbations. In particular, the points in the $M(R)$ or $M(\rho_c)$ relations where eigenfrequencies are null are not their critical points anymore. This is expected because the classical rules of dynamical stability of stars assume perfect fluids \citep{Wheeler1965,1986bhwd.book.....S,1988ApJ...325..722F}. 
The fact that the terminal central density (where $\omega=0$) of an elastic star is larger than those of a perfect fluid is also reasonable, because elastic shear stresses increase the pressure of the system, allowing for denser objects. This is in agreement with the more general analysis of \cite{2019PhRvD..99j4072R}, who showed that even ultracompact stars are possible when larger anisotropic pressures take place. 
An interesting consequence for systems with $\partial M/\partial \rho_c<0$ is that they could be more compact, which could lead to a richer phenomenology that may be probed with electromagnetic missions and relativistic ray-tracing models (see, e.g., \cite{2018ApJ...855..116V,Raaijmakers_2019}).

We have shown that the radial stability of hybrid stars strongly depends on the value of the shear modulus of the elastic quark phase, which at present is not yet known. The key point is whether or not stability changes due to it could be detected. In terms of GW measurements, the hope for very precise radius constraints (inferred from the component masses and tidal deformability measurements) lies basically in the third generation detectors, or in collecting many detections which will lead to analyses unraveling the NS interior through features visible in the population of measurements, or in a lucky nearby GW170817-like strong signal, etc. Therefore, it is an issue in principle for the future. 

The alternative relies on electromagnetic observations. The NICER mission, which is currently collecting data from some compact systems, has already delivered NS radius constraints with $10$\% accuracy. For stars which are brighter, have favourable geometries (e.g., particular inclination angles with respect to the observer and colatitudes of the polar caps), etc.,  it might be possible that even smaller uncertainties are reached, for instance breaking the $5\%$ threshold \citep{2016ApJ...832...92O}. Assuming that there is a quark phase in the interior of stars, our estimates suggest that NICER might only be able to constrain (probably unrealistically) large quark shear moduli. Future missions, such as the eXTP and ATHENA, might be more promising because they will be able to measure NS radii more accurately.

However, one should also bear in mind that the maximum increase of radius differences for stable stars due to elasticity is EOS-dependent. Naturally, this brings ambiguities to the shear modulus extraction when the EOS (and associated radius) uncertainties are large enough, which is currently the case.  \footnote{State-of-the-art many-body perturbative calculations with two and three nucleon interactions up to the next-to-next-to-next-leading order for cEFT suggest that for a density around the saturation density, the fractional pressure uncertainty is around $10-20\%$ \citep{2020PhRvL.125t2702D}. For smaller densities, the uncertainties are smaller. (For larger densities, uncertainties based on cEFT increase but they are not reliable because its perturbative scheme breaks down. Phenomenological potentials for nucleon-nucleon interactions might be more appropriate in this case \citep{2019arXiv190311353B}.) Roughly speaking, this would imply relative uncertainties of the radius of a $1.4M_{\odot}$ neutron star around $15\%$ \citep{2020Sci...370.1450D,2020PhRvC.102e5803E}. These uncertainties only come from microscopic models. They decrease to around $5-10\%$ when both a maximum mass to stars and GW and current electromagnetic observations are taken into account \citep{2020NatAs...4..625C,2020Sci...370.1450D,2020PhRvC.102e5803E}. More observations (or higher precision for some observations) could decrease the above uncertainties. For the quark phase's EOS, the uncertainties there are model-dependent and mostly determined by means of astrophysical observations (maximum masses and GWs) \citep{Annala:2019puf}. They could vary in a wide range ($\sim 10-50\%$) depending on the maximum speed of sound of the quark phase \citep{Annala:2019puf}. Radius uncertainties of hybrid stars depend on aspects of the phase transition. If the transition happens at low masses, current observations lead to larger radius uncertainties for hybrid stars than purely hadronic stars because the former could be more compact \citep{2018PhRvL.120z1103M}.} A possible way to observationally address this problem is to have several radius measurements of stars with similar masses and attempt to constrain the EOS and the shear modulus simultaneously.  The required number of measurements depends on the parameters of the EOS and the shear modulus. In the simplest case it would account for an extra degree of freedom and it would require just an additional observation with respect to approaches for hybrid EOS constraints. The mass value on which to focus should be, broadly speaking, between $M^{\rm elas,s}_{\omega=0}$ and $M_{\rm max}$ (for a given EOS and shear modulus). Since the minimum value for $M^{\rm elas,s}_{\omega=0}$ is not known, no mass should be disregarded in principle; however, our analysis roughly suggests that stars with masses larger than approximately $(1.7-1.8)\,M_{\odot}$ might be interesting candidates for probing the elasticity of the quark phase.
It is still to be checked if such stars could be feasible candidates for the NICER, eXTP and ATHENA missions.

In the case of unstrained backgrounds, one would expect that the difference between the terminal radius (for which $\omega^2=0$) of a hybrid star with quark elasticity and another one without it to be weakly sensitive to EOS uncertainties.
This could be roughly seen from the perturbation equations. Since the terminal radii are related to the fundamental mode, $\xi$ and $\Delta p$ would be slowly varying functions, especially so in the quark phase. In this case and from Eqs. \eqref{xi_bar_eq} and \eqref{tov1}, one sees that $[\Delta p/(p_0\Gamma) + 3\bar{\xi}]\sim -rp_0'/(p_0+\rho_0)\xi(0)= r\bar{\xi}(0)\nu_0'/2 \sim \bar{\xi}(0) m/r$. Therefore, the only meaningful component of the shear stress, Eq. \eqref{Delta_p11_final}, would be only weakly dependent on pressure uncertainties. 
We have checked numerically that indeed the difference between the terminal radii changes by a fraction of a percent when  the quark parameters of our book-keeping hybrid EOS H2 are changed. As a result, one might expect a somewhat characteristic dispersion of radii for a mass range exclusively due to the elasticity of the quark phase. An important improvement in our approach would be considering elasticity already at the background level, given that the shear modulus for some mass ranges might be a non-negligible fraction of the pressure. We leave this to be carried out elsewhere.

Just for completeness, we stress that all current observational constraints to NSs are also in agreement with purely hadronic EOSs \citep{Annala:2019puf,2020Sci...370.1450D}. In this case, elastic aspects would play a negligible role in the stability of NSs because they would just appear at relatively low densities (below the nuclear saturation density), encompassing only a fraction of the total mass of the star. The elasticity contribution to the stability of a neutron star would just become relevant if it appeared at high densities, such as in the quark phase. That is why we have taken it as a working assumption. Only when radius uncertainties are a few percent---a possibility with future missions such as eXTP and ATHENA---we will be able to constrain EOSs to very high accuracy and thus directly probe an elastic quark phase in stars.

The elasticity of the quark phase could also leave an important imprint on the stellar eigenfrequencies. We have found that the relative changes (with respect to the perfect-fluid case) are larger for stiff EOSs, and they could be larger than $10\%$ ($\sim 20 \%$ for the squared frequencies---related to the energies of the modes) for the massive stars to be probed. Hence, if NS radial modes could be observed, the quark elasticity might in principle be probed. A possible way of doing that would relate to modulations of the NS lightcurve \citep{2019ApJ...884L..16C}. However, this might work only for large enough amplitudes of oscillation, which are generally unknown. Another possibility concerns direct GW observations due to the coupling of radial modes with rotation \citep{1967ApJ...147..664C}. However, this seems a possibility only for postmerger stars \citep{2019A&A...622A.194D}, where our analysis is not reliable due to the use of cold EOS candidates and the unknown impact rapid rotation could have on the terminal mass. In this case, though, matching conditions related to rapid conversions seem more appropriate; we leave these analyses for future work.

Finally, we point out that depending on the nature of the EOS, a large extended branch of (metastable) NSs could appear if conversions are slow. In this case in particular, the terminal mass (where $\omega=0$) could be significantly different from the maximum mass. As a result, it might reach the range of masses more commonly observed for NSs. 
One question in this context is whether this is relevant for probing an elastic quark phase. 
Our analysis suggests that it seems unlikely to probe the elasticity of the quark phase electromagnetically if the EOS is soft or intermediate. The possibility might exist for stiff EOSs, but satellite missions such as eXTP and ATHENA might be required.

Summing up, in this work we have laid out the problem of radial stability for hybrid stars with elastic quark phases and have shown that their mass-radius region of dynamical stability is extended with respect to their perfect-fluid counterpart. 
Our analysis suggests that significant increases of radius differences between elastic and perfect-fluid stars with null frequencies might take place when the shear modulus of the quark phase is roughly larger than 50 MeV fm$^{-3}$.

\section{Acknowledgements}

We thank the anonymous referee for constructive comments which have improved our work. We thank Andreas Schmitt and Nils Andersson for helpful discussions in an earlier version of this work. J.P.P. and M.B. acknowledge the financial support from the Polish National Science Centre grant No. 2016/22/E/ST9/00037. J.P.P. is also thankful for partial support at an earlier stage of this paper given by Funda\c c\~ao de Amparo \`a Pesquisa do Estado de S\~ao Paulo (FAPESP) under grants No. 2015/04174-9 and No. 2017/21384-2. L.T. acknowledges FAPESP for support under grants No. 2018/04281-8 and No. 2019/15124-3. G.L. is thankful to the Brazilian agency Conselho Nacional de Desenvolvimento Cient\'{\i}fico e Tecnol\'ogico (CNPq) for financial support.
M.S. acknowledges support of the TEAM/2016-3/19 grant from FNP. P.H. and J.L.Z. were supported by the National Science Centre, Poland grant 
2018/29/B/ST9/02013.

\bibliographystyle{aasjournal}
\bibliography{crystalline}

\end{document}